\documentstyle[11pt,adassconf]{article}  

\begin{document}   

\paperID{P1-39}

\title{Immersive 4D Interactive Visualization of Large-Scale Simulations}

\author{Peter Teuben}
\affil{Astronomy Department, University of Maryland, College Park, MD}
\author{Piet Hut}
\affil{Institute for Advanced Study, Princeton, NJ}
\author{Stuart Levy}
\affil{National Center for Supercomputing Applications, 
        University of Illinois Urbana-Champaign, Urbana, IL}
\author{Jun Makino}
\affil{Department of Astronomy,
  The University of Tokyo,
  Bunkyo-ku,
  Tokyo 113-0033,
  JAPAN.}
\author{Steve McMillan}
\affil{ Department of Physics and Atmospheric Science
         Drexel University, Philadelphia, PA}
\author{Simon Portegies Zwart\altaffilmark{1}}
\affil{Massachusetts Institute of Technology, Cambridge, MA}
\author{Mike Shara,  Carter Emmart}
\affil{American Museum of Natural History, New York, NY}
\altaffiltext{1}{Hubble Fellow}

\contact{Peter Teuben}
\email{teuben@astro.umd.edu}

\paindex{Teuben, P.J.}
\aindex{Hut, P.}
\aindex{Levy, S.}
\aindex{Makino, J.}
\aindex{McMillan, S.}
\aindex{Portegies Zwart, S.}
\aindex{Shara, M.}
\aindex{Emmart, C.}

%
%


\begin{abstract}          

In dense clusters a bewildering variety of interactions between stars can be
observed, ranging from simple encounters to collisions and other
mass-transfer encounters. With faster and special-purpose computers like
GRAPE, the amount of data per simulation
is now exceeding 1TB. Visualization of such data
has now become a complex 4D data-mining problem, combining space and time,
and finding interesting events in these large datasets. We have recently
starting using the virtual reality simulator, installed in the Hayden
Planetarium in the American Museum for Natural History, to tackle some of
these problem. \htmladdnormallinkfoot{This work}{http://www.astro.umd.edu/nemo/amnh/} 
reports on our first ``observations'',
modifications needed for our specific experiments, and perhaps field ideas
for other fields in science which can benefit from such immersion. We also
discuss how our normal analysis programs can be interfaced with this kind of
visualization.

\end{abstract}

\section{NEMO, Starlab and GRAPE}

\htmladdnormallinkfoot{NEMO}{http://www.astro.umd.edu/nemo/} 
(Teuben 1994) and
\htmladdnormallinkfoot{Starlab}{http://www.manybody.org}
are traditional programming environments with which
N-body simulations can be setup, run and analyzed. NEMO also has
a number of tools to import and export data in tables, CCD type images,
FITS files and a large number of other N-body formats. NEMO is more
geared towards collisionless stellar dynamics, while Starlab has
more sophisticated programs to deal with close encounters, and can
now also incorporate stellar evolution through the 
SEBA package (Portegies Zwart et al. 2001).
NEMO and Starlab present themselves to a user as a large set of programs,
often glued together using pipes in shell scripts to set up and
run complex simulations. For the programmer, a large set of classes and
functions are available to construct new integrators and analysis programs.
For example, in the following Starlab 
example an anisotropic King model with 2048 
particles has been evolved with 50\% binaries (i.e. 3096 actual stars)
and stellar evolution:

\footnotesize
\begin{verbatim}

mk_aniso_king -i -n 2048 -u -w 4 -F 3                      |\
  mkmass -i -u 100 -l 0.1 -f 3                             |\
  mksecondary -f 0.5 -l 0.1                                |\
  addstar -Q 0.5 -R 2.5                                    |\
  scale -M 1 -E -0.25 -Q 0.5                               |\
  mkbinary -f 2 -l 1 -u 1000000 -o 2                       |\
  kira -a 0.1 -d 1 -D 25 -n 25 -t 4000 -Q -G 2 -u -B -z 1 > run001

\end{verbatim}
\normalsize

The GRAPE special purpose hardware (Hut and Makino, 1999),
now running at 100 TerraFlops speed, has
been successfully interfaced with Starlab, and now is starting to
produce massive datasets. Analysis and visualization 
techniques of those dataset are becoming increasingly
challenging.

\section{AMNH, Virtual Director and Partiview}

The American Museum for Natural History (AMNH) in New York City has
recently renovated its planetarium, and converted it into a
state-of-the-art digital planetarium with capabilities for scientific
visualization. Their computer system consists 
of an Onyx2, with 28 CPUs, 14 GB of memory, 2 TB diskspace and
7 graphics pipes. Each graphics pipe controls one of 7 
projectors which illuminate the dome in a dodecahedral pattern.
The software that drives most visualization is an NCSA product called
\htmladdnormallinkfoot{Virtual Director}
{http://virdir.ncsa.uiuc.edu/virdir/virdir.html} ({\tt virdir}),
that we have now
been using during a number of night sessions in the dome, much
like optical observers (during daytime the planetarium is of course
used for public viewing). It allows us to ``fly'' through
the data, in space and time.
By adding complete orbital information for a select number
of stars we have started fully interactive data mining of our
4D spacetime histories of these star cluster simulations runs.
\begin{figure}
\plotone{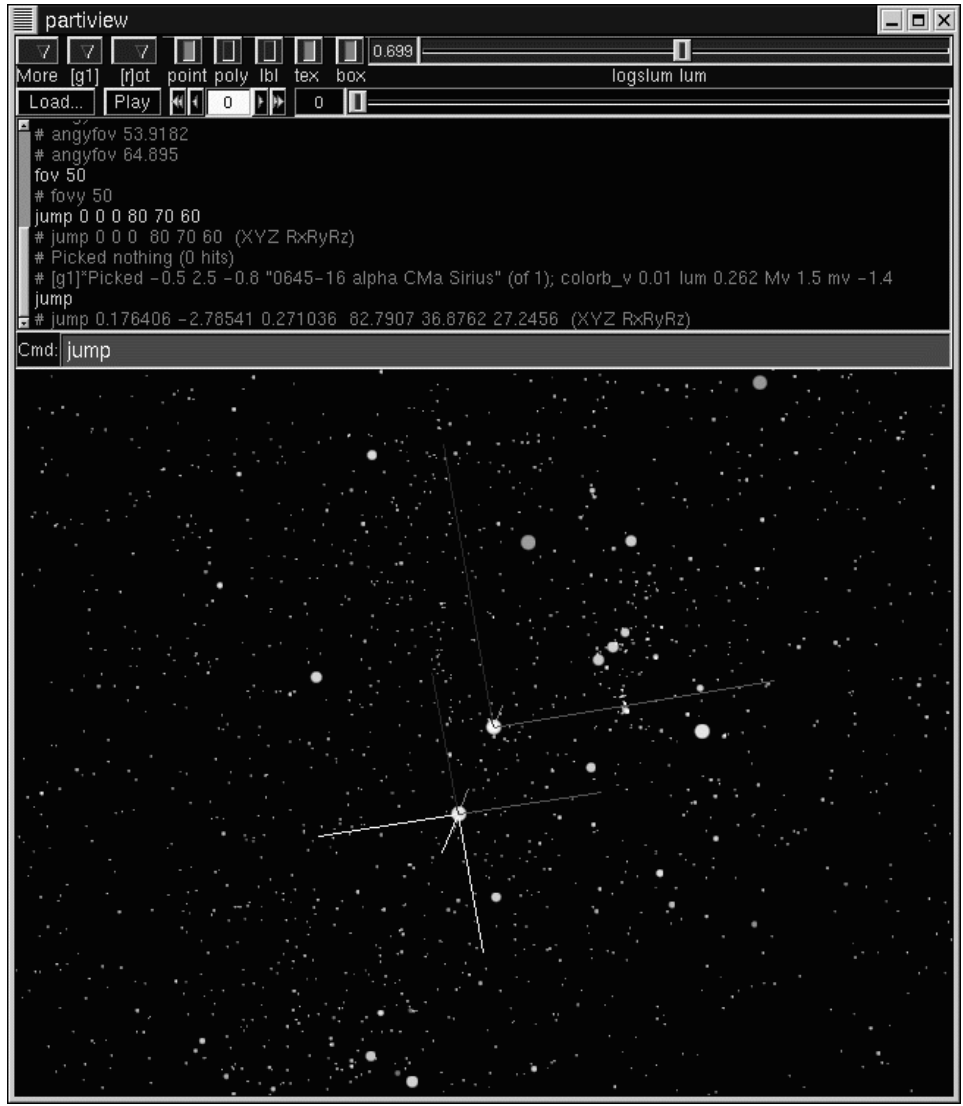}
\caption{Partiview in action: after loading a 4-D dataset, the mouse in
the window controls motion and spatial orientation, the 
jog-wheel at the top right controls animation, either manually or 
via the CD-like control buttons to
the right. Partiview also has a command language, commands
are entered in the {\tt Cmd} 
window in the middle,
right below the scrolling logfile window.
}
\label{P1-39-fig-1}
\end{figure}
In order for us to test new visualization techniques, 
algorithms and interfaces with the
Starlab  environment, we used an existing program
{\tt partiview}, which had been derived from {\tt virdir},
and which can be run on a normal workstation or laptop. It uses the
FLTK and MESA/OpenGL libraries for its user interface and
fast graphics. 
A screenshot of {\tt partiview} in action 
can be seen in Figure~\ref{P1-39-fig-1}.  We have modified {\tt partiview}
to understand our Starlab simulatation data, and added interfaces that allow
this workstation version to animate and move in time and space. 
{\tt partiview} comes with a small but powerful set of commands with which
dataselections and viewing can be made, and we hope to expand this into a more
mature scripting language. It is also fairly straightforward 
for other packages to benefit from using {\tt partiview}.

\section{Future Plans}

In the spirit of the federated model of archiving observational data,
recently proposed by the NVO (National Virtual Observatory) initiative,
we will develop a Starlab-based archive.  A simulation of a globular
cluster with a million stars stars for ten billion years will generate
100 Tbytes of raw data, of which we would like to store at least 1
Tbyte, and preferably more, for 4D visualization of the full history of
the evolution of a star cluster.  Although our main goal will be to
enable rapid and intelligent access to our simulation output files, we
will simultaneously develop a flexible and transparent interface with
the NVO databace and protocols.  Our Starlab policy will be to make
all simulation results freely and publicly available to `guest observers'.

\acknowledgements{Part of this paper was written while we were
visiting the American Museum of Natural History.  We acknowledge 
the hospitality of their
astrophysics department and visualization group.  We thank the
Alfred P. Sloan Foundation for a grant to Hut for observing
astrophysical computer simulations in the Hayden Planetarium at the
Museum. NCSA's Virtual Director group comprises of Donna Cox, Robert
Patterson and Stuart Levy.
}

\end{document}